\documentstyle[12pt]{article}

\newcommand{\bce}{\begin{center}}
\newcommand{\ece}{\end{center}}
\newcommand{\beq}{\begin{equation}}
\newcommand{\eeq}{\end{equation}}
\newcommand{\bea}{\vspace{0.25cm}\begin{eqnarray}}
\newcommand{\eea}{\end{eqnarray}}

\newcommand{\ba}{\begin{array}}
\newcommand{\ea}{\end{array}}

\def\lsim{\mathrel{\rlap{\lower4pt\hbox{\hskip1pt$\sim$}}
    \raise1pt\hbox{$<$}}}         
\def\gsim{\mathrel{\rlap{\lower4pt\hbox{\hskip1pt$\sim$}}
    \raise1pt\hbox{$>$}}}         

\def\Pom{{\bf I\!P}}

\def\lsim{\mathrel{\rlap{\lower4pt\hbox{\hskip1pt$\sim$}}
    \raise1pt\hbox{$<$}}}         
\def\gsim{\mathrel{\rlap{\lower4pt\hbox{\hskip1pt$\sim$}}
    \raise1pt\hbox{$>$}}}         

\def\Pom{{\bf I\!P}}

 \advance\oddsidemargin  by -0.9in
  \advance\evensidemargin by -0.9in
  \advance\topmargin      by -0.20in
\makeindex
\begin{document}

\phantom{.}\hspace{10.5cm}{\large \bf 16 August 2000}
\vspace{1.5cm}\\
\begin{center}{ \Large \bf
Color Dipole Systematics of the Diffraction Slope \\
in Diffractive Photo- and Electroproduction \\
\vspace*{0.2cm}
of Vector Mesons}
\vspace*{2.5cm} \\
{\large \bf J.~Nemchik}
\vspace*{0.3cm} \\
{\it Institute of Experimental Physics, Slovak Academy of Sciences,\\
Watsonova 47, 04353 Ko\v sice, Slovakia} \\
\vspace*{0.5cm}
\vspace*{5.0cm}
\begin{minipage}[h]{13cm}
\centerline{\Large \bf
Abstract }
\vspace*{0.5cm}

We present the first evaluation of the color dipole 
diffraction slope
from the data on diffractive photo- and electroproduction
of vector mesons. The energy and dipole size dependence
of the found dipole diffraction slope are consistent with
the color dipole gBFKL dynamics.
\end{minipage}
\end{center}
\pagebreak
\setlength{\baselineskip}{0.515cm}

%
%

Diffractive photo- and electroproduction of 
vector mesons, 
%
%
\beq
\gamma^{*}p \rightarrow Vp\,~~,  ~~~~~ V = \rho^{0}, \omega^{0},
\phi^{0}, J/\Psi, \Upsilon,...   
\label{eq:0}
\eeq
%
%
studied within the color dipole model of high energy scattering
\cite{KZ91,NNN92,KNNZ93}
(see also \cite{NZ91,NZ94,NZZ94}) 
offers an unique possibility to
scan the color dipole cross section \cite{KZ91,NNN92,KNNZ93,NNZscan}, which 
represents a fundamental quantity and reflects
the interaction of the relativistic color dipole of the dipole
moment (size) ${\bf r}$ with the target nucleon.
The alternative description of vector meson production in terms of the
gluon structure function of the proton is presented in \cite{Ryskin,Brodsky}.
Because of the shrinkage of the transverse size of
the virtual photon with virtuality $Q^{2}$, the
vector meson production amplitude scans the color
dipole cross section at the dipole size  $r\sim  
r_{S}$, where the scanning radius $r_{S}$ 
can be expressed through the scale parameter $A$
\cite{KNNZ94,NNZscan}
%
\beq
r_{S} \approx {A \over \sqrt{m_{V}^{2}+Q^{2}}}\, ,
\label{eq:1}
\eeq
%
where 
$m_{V}$ is the vector meson mass and $A\approx 6$.
This scanning phenomenon can be also applied
for the dipole diffraction slope leading to a decrease
of the slope with $Q^{2}$ \cite{NZZslope,NNPZZ96bfkl}  supported
by the available experimental data (see below).
Detailed analysis of the $Q^{2}$ dependence of the Regge
growth of the diffration slope for production of
charmonium and bottonium states has been presented
in the paper \cite{NNPZZ96bfkl}.
Specifically, the leading twist terms of production
amplitudes of the transversely (T) and longitudinally (L)
polarized 1S vector mesons in the
expansion over the relevant short-distance parameter
$r_{S}^{2} \propto 1/(Q^{2}+m_{V}^{2})$ are of the form
\cite{KNNZ94,NNZscan}
%
%
\beq
{\rm Im}
{\cal M}_{T} \propto
{1 \over Q^2+m_{V}^{2}}\sigma(x_{eff},r_{S}) \propto
r_{S}^{2}\sigma(x_{eff},r_{S})\, ,
\label{eq:2a}
\eeq
%
%
%
%
\beq
{\rm Im}
{\cal M}_{L} \approx {\sqrt{Q^{2}}\over m_{V}}{\cal M}_{T} \propto
{\sqrt{Q^{2}}\over m_{V}}r_{S}^{2}
\sigma(x_{eff},r_{S})
\, .
\label{eq:2b}
\eeq
%
As the result the dipole cross section can be exctracted
from the vector meson production cross section.
The first evaluation of the dipole cross section 
from the data on diffractive photo-
and electroproduction of vector mesons has
been presented in the paper \cite{NNPZdipole}.
Following the generalization of the color dipole
factorization formula for vector meson production
amplitude \cite{KZ91,NNN92,KNNZ93,NNPZ97}
%
%
\beq
{\rm Im} {\cal M}(\gamma^{*}\rightarrow V,x_{eff},Q^{2})=
\int\limits_{0}^{1} dz\int d^{2}{\bf {r}}
\sigma(x_{eff},r)
\Psi_{V}^{*}(r,z)\Psi_{\gamma^{*}}(r,z)\, 
\label{eq:3}
\eeq
%
%
to the diffraction slope of the reaction
$\gamma^{*}p\rightarrow Vp$ one can write \cite{NNPZZ96bfkl}
%
%
\bea
B(\gamma^{*}\rightarrow V,x_{eff},Q^{2})
{\rm Im} {\cal M}(\gamma^{*}\rightarrow V,x_{eff},Q^{2})=
\nonumber \\
\int\limits_{0}^{1} dz\int d^{2}{\bf r}
B(x_{eff},r)\sigma(x_{eff},r)
\Psi_{V}^{*}(r,z)\Psi_{\gamma^{*}}(r,z)\, ,
\label{eq:4}
\eea
%
%
and one can extract also 
the dipole diffraction slope $B(x_{eff},r)$
from the vector meson production cross sections using
analogical procedure as that for
the color dipole cross section extraction
published in \cite{NNPZdipole}.

In this paper we present the results of the
first evaluation of the dipole diffraction slope from the
data on real photoproduction and electroproduction of
vector mesons ($\rho^{0}$, $\phi^{0}$, and $J/\Psi$)
from the fixed target and collider HERA experiments.
We find color blindness of the dipole diffraction cone similarly
to that  for the color dipole cross section. We verify
approximate flavor independence  of the $B(x_{eff},r)$ in
the scaling variable $Q^{2}+m_{V}^{2}$  leading to the close
values of the diffraction slope when one  takes  the same
values of the scanning radius.
The data from HERA experiments allow also the first evaluation
of energy dependence of the dipole diffraction cone at different
dipole sizes and confirm the predictions from the BFKL 
dynamics that a shrinkage with energy of the slope parameter is
larger at larger dipole sizes.


We start with the probability amplitudes to find color
dipole of size ${\bf r}$  in the photon and vector meson.
The details of calculation of the diffractive amplitudes have been
presented elsewhere \cite{NNZscan,NNPZ97,NNPZZ96bfkl}. For the $Vq\bar{q}$
vertex function we assume the Lorentz structure $\Gamma\bar{\Psi}
\gamma_{\mu}\Psi V_{\mu}$.
For the $s$-channel helicity conservation  at small ${\bf q}$,
transverse (T) photons produce the transversely polarized vector
mesons and the longitudinally polarized (L) photons (to be more
precise, scalar photons) produce longitudinally polarized
vector mesons. One finds
%
%
\bea
{\rm Im}{\cal M}_{T}(\gamma^{*}\rightarrow V,x_{eff},Q^{2})=
{N_{c}C_{V}\sqrt{4\pi\alpha_{em}} \over (2\pi)^{2}}
\cdot~~~~~~~~~~~~~~~~~~~~~~~~~~~~~~~~~
\nonumber \\
\cdot \int d^{2}{\bf{r}} \sigma(x_{eff},r)
\int_{0}^{1}{dz \over z(1-z)}\left\{
m_{q}^{2}
K_{0}(\varepsilon r)
\phi(r,z)-
[z^{2}+(1-z)^{2}]\varepsilon K_{1}(\varepsilon r)\partial_{r}
\phi(r,z)\right\} \nonumber \\
= {C_{V} \over (m_{V}^{2}+Q^{2})^{2}}
\int {dr^{2} \over r^{2}} {\sigma(x_{eff},r) \over r^{2}}
W_{T}(Q^{2},r^{2}) \nonumber \\
= g_{T}\sqrt{4\pi\alpha_{em}}~C_{V}\sigma(x_{eff},r_{S})
\frac{m_{V}^{2}}{m_{V}^{2}+Q^{2}}
\label{eq:5}
\eea
%
%
%
%
\bea
{\rm Im}{\cal M}_{L}(\gamma^{*}\rightarrow V,x_{eff},Q^{2})=
{N_{c}C_{V}\sqrt{4\pi\alpha_{em}} \over (2\pi)^{2}}
{2\sqrt{Q^{2}} \over m_{V}}
\cdot~~~~~~~~~~~~~~~~~~~~~~~~~~~~~~~~~
 \nonumber \\
\cdot \int d^{2}{\bf{r}} \sigma(x_{eff},r)
\int_{0}^{1}dz \,K_{0}(\varepsilon r)\left\{
[m_{q}^{2}+z(1-z)m_{V}^{2}]
\phi(r,z)-\partial_{r}^{2}
\phi(r,z)\right\} \nonumber \\
= {C_{V} \over (m_{V}^{2}+Q^{2})^{2}}
{2\sqrt{Q^{2}} \over m_{V}}
\int {dr^{2} \over r^{2}} {\sigma(x_{eff},r) \over r^{2}}
W_{L}(Q^{2},r^{2}) \nonumber \\
= g_{L}\sqrt{4\pi\alpha_{em}}~C_{V}\sigma(x_{eff},r_{S})
\frac{\sqrt{Q^{2}}}{m_{V}}\frac{m_{V}^{2}}{m_{V}^{2}+Q^{2}}
\label{eq:6}
\eea
%
%
where
%
%
\beq
\varepsilon^{2} = m_{q}^{2}+z(1-z)Q^{2}\,,
\label{eq:7}
\eeq
%
%
$\alpha_{em}$ is the fine structure constant, $N_{c}=3$ is the number
of colors, $C_{V}={1\over \sqrt{2}},\,{1\over 3\sqrt{2}},\,{1\over 3},\,
{2\over 3},\,{1\over 3}~~$ are the charge-isospin factors for the
$\rho^{0},\,\omega^{0},\,\phi^{0},\, J/\Psi, \Upsilon$ production,
respectively and $K_{0,1}(x)$ are the modified Bessel functions.
The detailed discussion and parameterization of the lightcone radial
wave function $\phi(r,z)$ of the $q\bar{q}$ Fock state of the vector
meson is given in \cite{NNPZ97}. 
The terms $\propto
K_{0}(\varepsilon r)\phi(r,z)$ and
$\propto \varepsilon K_{1}(\varepsilon r)\partial_{r}\phi(r,z)$
for (T),
$K_{0}(\varepsilon r)\partial_{r}^{2}\phi(r,z)$ for (L)
correspond to the helicity conserving and helicity-flip
transitions in the $\gamma^{*}\rightarrow q\bar{q},
V\rightarrow q\bar{q}$ vertices, respectively.

In (\ref{eq:5}), (\ref{eq:6})
the energy dependence of the dipole cross section is quantified
in terms of an effective value of the Bjorken variable, $x_{eff}$,
which is connected with dimensionless 
rapidity, $\xi=\log{1\over x_{eff}}$ and reads
%
%
\beq
x_{eff} = \frac{m_{V}^{2}+Q^{2}}{2\nu m_{p}}
\sim \frac {Q^{2}+m_{V}^{2}}{W^{2}}\, ,
\label{eq:8}
\eeq
%
%
where $m_{p}$ and $m_{V}$ is the proton mass and mass of
vector meson, respectively.
Hereafter, we will write the energy dependence of the dipole
cross section in both variables,
either in $\xi$ or in $x_{eff}$.

Normalization of production  amplitudes 
(\ref{eq:5}), (\ref{eq:6}) satisfies the following relation
%
%
\beq
\frac{d\sigma}{dt}|_{t=0} = \frac{|{\cal M}|^{2}}{16\pi}
\label{eq:9}
\eeq
%

A small real part of  production amplitudes can be taken
in the form \cite{GribMig}
%
%
\beq
{\rm Re}{\cal M}(\xi,r) =\frac{\pi}{2}\cdot\frac{\partial}
{\partial\xi} {\rm Im}{\cal M}(\xi,r)\,.
\label{eq:9a}
\eeq
%
%
and can be easily included in the production amplitudes
using substitution
%
%
\beq
\sigma(x_{eff},r)\rightarrow 
\biggl (1-i\frac{\pi}{2}\frac{\partial}{\partial~log~x_{eff}}
\biggr)
\sigma(x_{eff},r) = \biggl [1-i\alpha_{V}(x_{eff},r)
\biggr  ]\sigma(x_{eff},r)
\label{eq:10}
\eeq

Within the mixed $({\bf{r}},z)$ representation,
the high energy meson is considered as
a system of color dipole described by
the distribution
of the transverse separation ${\bf{r}}$ of the quark and
antiquark given by the $q\bar{q}$ wave function,
$\Psi({\bf{r}},z)$, where $z$ is
the fraction of meson's lightcone momentum
carried by a quark.
The Fock state expansion for the
relativistic meson starts
with the $q\bar{q}$ state and
the higher Fock states $q\bar{q}g...$
become very important at high energy $\nu$.
The interaction of the relativistic
color dipole of the dipole moment, ${\bf{r}}$, with the
target nucleon is quantified by the energy dependent color
dipole cross section, $\sigma(\xi,r)$,
satisfying
the gBFKL equation
\cite{NZ94,NZZ94} for the energy evolution.
This reflects the fact that 
in the leading-log ${1\over x}$ approximation the
effect of higher Fock states can be
reabsorbed into the energy dependence
of $\sigma(\xi,r)$.
The dipole cross section is flavor
independent and represents the universal
function of $r$ which describes
various diffractive processes in unified form.
At high energy, when the transverse separation, ${\bf{r}}$,
of the quark and antiquark is frozen during the interaction
process, 
the scattering
matrix describing the $q\bar{q}$-nucleon interaction
becomes diagonal
in the mixed $({\bf{r}},z)$-representation ($z$ is known also as
the Sudakov light cone variable). 
This diagonalization property is held even 
when the dipole size, ${\bf{r}}$, is large,
i.e. beyond the perturbative region of short distances.
Color  dipole factorized form of formulas
(\ref{eq:5}), (\ref{eq:6})
follows from such diagonalization property.
The detailed discussion about the space-time
pattern of diffractive electroproduction of vector mesons
is presented in \cite{NNPZ97,NNPZZ96bfkl}.

Eqs. (\ref{eq:5}), (\ref{eq:6}) are related to the pure
pomeron exchange, which is important at large values
of parameter $\omega=1/x_{eff}$. However, at moderate
and small values of $\omega$ there is a substantial
contribution to the $\gamma N $ total cross section
comming from the non-vacuum Reggeon exchange.
The Regge fit to the $\gamma  p$ total cross section can be
cast in the following form \cite{DL}
%
%
\beq
\sigma_{tot}=\sigma_{\Pom}(\gamma p)\biggl  (1+\frac{B}{\omega^{\Delta}}
\biggr )
\label{eq:11}
\eeq
%
%
where $B=2.332$ and $\delta=0.533$ according to Donnachie-Landshoff
fit. In (\ref{eq:11}) the term $B/\omega^{\Delta}$ in
the factor $f=1+B/\omega^{\Delta}$ represents the non-vacuum
Reggeon exchange contribution, which is similar also in
real $\rho^{0}$ photoproduction amplitude.
At large $Q^{2}$ we assume that the Reggeon/pomeron ratio
scales with $\omega$. It  is consistent with the known
decomposition of the proton structure function into the valence
(non-vacuum Reggeon) and sea (pomeron) contributions.
Then, for example, at NMC energy $f=1.25$ at $\omega\sim 70$
relevant to $Q^{2}= 3 ~GeV^{2}$ and
$f=1.80$ at $\omega\sim 9$
relevant to $Q^{2}= 20 ~GeV^{2}$. At energy attainable at HERA
the non-vacuum Reggeon exchange contribution can be neglected
because of a large value of the Regge parameter $\omega$.
For other vector mesons ($\phi^{0}, J/\Psi, \Upsilon$) one
expects $f=1$ due to the Zweig rule.
Thus, the forward differential cross section for $\rho^{0}$
photo- and electroproduction reads:
%
%
\beq
\frac{d\sigma(\gamma^{*}\rightarrow V)}{dt}|_{t=0}=
f^2\frac{d\sigma_{\Pom}(\gamma^{*}\rightarrow V)}{dt}|_{t=0}
\label{eq:12}
\eeq
%
%
In (\ref{eq:5}) and (\ref{eq:6}) we separate out  the
scanning radius $r_{S}$ and then the so introduced coefficient
functions $g_{T,L}$ are smooth functions of $Q^{2}$.
Such a procedure is detaily described in the paper \cite{NNPZdipole}.

Taking into account the contribution of the real part 
and the non-vacuum Reggeon exchange 
to the production amplitude then the experimentally measured forward cross
section reads
%
%
\bea
\frac{d\sigma(\gamma^{*}\rightarrow V)}{dt}|_{t=0}=
\frac{f^{2}}{16\pi}\biggl [(1+\alpha_{V,T}^{2}){\cal M}_{T}^{2}+
\epsilon(1+\alpha_{V,L}^{2}){\cal M}_{L}^{2}\biggr ]  \nonumber \\
=
\frac{f^{2}}{16\pi}(1+\alpha_{V}^{2})\biggl [{\cal M}_{T}^{2}
+\epsilon{\cal M}_{L}^{2}\biggr ]\,\,  ,
\label{eq:13}
\eea
%
%
neglecting the difference between $\alpha_{V,T}$ and $\alpha_{V,L}$
for the transverse and longitudinal cross sections, respectively.
Then one can exctract the color dipole cross section
from the data on forward production cross section using
the above determined coefficient functions $g_{T}$ and
$g_{L}$ and Eqs. (\ref{eq:5}), (\ref{eq:6}), (\ref{eq:9}) 
and (\ref{eq:13}).
%
%
\bea
\sigma(x_{eff},r_{S})=\frac{1}{f}\frac{1}{C_{V}}
\frac{Q^{2}+m_{V}^{2}}{m_{V}^{2}}\frac{2}{\sqrt{\alpha_{em}}}
\nonumber \\
\times \biggl (g_{T}^{2}+\epsilon\frac{Q^{2}}{m_{V}^{2}}g_{L}^{2}
\biggr )^{-1/2}~(1+\alpha_{V}^{2})^{-1/2}\sqrt{\frac{d\sigma(\gamma^{*}
\rightarrow V)}{dt}|_{t=0}}\, ,
\label{eq:14}
\eea
%
%
where $\epsilon$ is the longitudinal polarization of  the photon.
Very often the data are presented in the form of the $t$- integrated
production cross section $\sigma_{tot}(\gamma^{*}\rightarrow V)$. 
In this case we evaluate 
%
%
\beq
\frac{d\sigma(\gamma^{*}\rightarrow V)}{dt}|_{t=0}=
B(\gamma{*}\rightarrow V,x_{eff},Q^{2})
\sigma_{tot}(\gamma^{*}\rightarrow V)
\label{eq:15}
\eeq
%
%
taking the diffraction slope from the same publication.

Now the generalization of the above procedure for extraction
of the dipole diffraction slope from the data is following.
Comming from the matrix element (\ref{eq:4}) and taking into account
the scanning property one can conclude
that the main contribution to the amplidute comes
from the dipole size $r_{B}\sim 5/3~r_{S}$ because 
of $r^{2}$ behaviour of the diffraction  slope within
the color dipole gBFKL dynamics \cite{NZZslope,NNPZZ96bfkl}.
Then one can introduce another coefficient functions
$h_{T}$ and $h_{L}$:
%
%
\bea
B(\gamma^{*}\rightarrow V,x_{eff},Q^{2}){\cal M}_{T}(x_{eff},Q^{2})
= \nonumber \\
{C_{V} \over (m_{V}^{2}+Q^{2})^{2}}
\int {dr^{2} \over r^{2}} {\sigma(x_{eff},r) \over r^{2}}
\frac{B_{T}(x_{eff},r)}{r^{2}}
{\tilde{W}}_{T}(Q^{2},r^{2}) = \nonumber \\
h_{T}\sqrt{4\pi\alpha_{em}}~C_{V}\sigma(x_{eff},r_{B})
B(x_{eff},r_{B})\frac{m_{V}^{2}}{m_{V}^{2}+Q^{2}}
\label{eq:16}
\eea
%
%
%
%
\bea
B(\gamma^{*}\rightarrow V,x_{eff},Q^{2}){\cal M}_{L}(x_{eff},Q^{2})
= \nonumber \\
{C_{V} \over (m_{V}^{2}+Q^{2})^{2}}
{2\sqrt{Q^{2}} \over m_{V}}
\int {dr^{2} \over r^{2}} {\sigma(x_{eff},r) \over r^{2}}
\frac{B_{L}(x_{eff},r)}{r^{2}}
{\tilde{W}}_{L}(Q^{2},r^{2}) = \nonumber \\
h_{L}\sqrt{4\pi\alpha_{em}}~C_{V}\sigma(x_{eff},r_{B})
B(x_{eff},r_{B})
\frac{\sqrt{Q^{2}}}{m_{V}}\frac{m_{V}^{2}}{m_{V}^{2}+Q^{2}}
\label{eq:17}
\eea
%
%
  \begin{figure}[tbh]
  \includegraphics{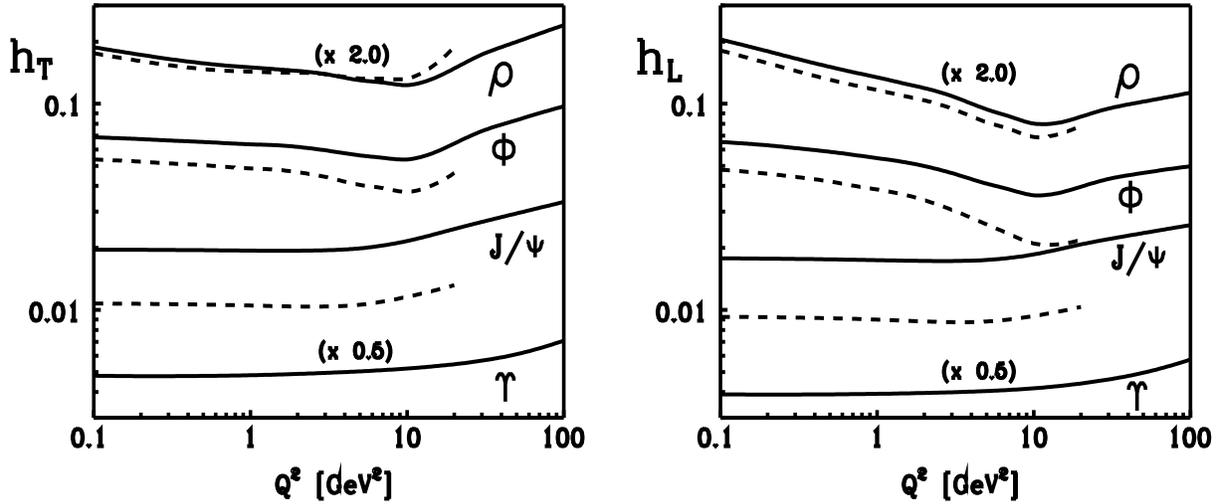}
  \begin{center}
  \vspace{6cm}
  \parbox{13cm}
  {\caption[Delta]
  {
    The $Q^{2}$ dependence of the
    coefficient functions $h_{T,L}$ at $W=15~GeV$
    (dashed curve) and $W=150~GeV$ (solid curve).
   }
  \label{event1}}
  \end{center}
  \end{figure}
%
%

The possibility of a such relationship between the production
amplitude multiplied by the diffraction slope for vector
meson production and
the dipole cross section multiplied by the 
dipole diffraction slope at a well
defined dipole size $r_{B}=\frac{5}{3}r_{S}$ 
has the same reasons as ones detaily
studied and discussed in the paper \cite{NNPZdipole}.
In Fig.1 we present the $Q^{2}$ dependence of $h_{T,L}$  for
different production processes at $W=15 ~GeV$ and $W=150 ~GeV$
reflecting the lower and upper limits of
the energy range from the fixed target to HERA
experiments. 
The residual smooth  $Q^{2}$ behaviour of the
coefficient functions $h_{T,L}$ is connected
with the smooth and well understood $Q^{2}$
dependence of the scale factors $A_{T,L}$, which are
presented in the rationship between the scanning radius
$r_{S}$ and the position $A_{T,L}/\sqrt{Q^{2}+m_{V}^{2}}$
of the peak of the functions ${\tilde{W}_{T,L}(Q^{2},r^{2})}$.
The detailed discussion about the similar smooth $Q^{2}$
dependence of the coefficient functions
$g_{T,L}$ is presented in \cite{NNPZdipole}.

Including the non-vacuum Reggeon contribution
and the contribution of the real part to the 
amplitude (\ref{eq:4})
the matrix element squared on r.h.s. of Eq.~(\ref{eq:4}) reads:
%
%
\bea
|{\cal M^{B}}(x_{eff},r)|^{2}=
|<\Psi_{V}(r,z)| \sigma(x_{eff},r)B(x_{eff},r) |\Psi_{\gamma^{*}}(r,z)>|^{2}
= \nonumber \\
f^{2}(1+\alpha_{V}^{2})4\pi\alpha_{em}C_{V}^{2}\frac{m_{V}^{4}}
{(m_{V}^{2}+Q^{2})^{2}}\sigma^{2}(x_{eff},r_{B})B^{2}(x_{eff},r_{B})
\biggl [h_{T}^{2}+\epsilon\frac{Q^{2}}{m_{V}^{2}}h_{L}^{2}\biggr ]
= \nonumber \\
B^{2}(\gamma^{*}\rightarrow V,x_{eff},Q^{2})|_{t=0}~16\pi
B(\gamma^{*}\rightarrow V,x_{eff},Q^{2})
\sigma_{tot}(\gamma^{*}\rightarrow V)\,\, ,
\label{eq:18}
\eea
%
%
where the last line is the l.h.s. squared of Eq. (\ref{eq:4}) representing
the multiplication of the forward diffraction slope squared 
for vector meson production and 
vector meson production amplitude squared, which can be expressed
via $t$- integrated  production cross section and the diffraction
slope using Eqs. (\ref{eq:9}) and (\ref{eq:15}).

The experimental determination of the forward 
diffraction cone $B(t=0)$ requires extrapolation of the differential
cross section $d\sigma/dt$ towards $t=0$, which is not always possible
and one often reports the $t$- integrated production cross sections
$\sigma_{tot}(\gamma^{*}\rightarrow V)$.
Following the high precision $\pi^{\pm}N$ scattering experiments,
the diffraction slope $B(t)$ depends  strongly on the region of $t$.
For the average $<t>~\sim 0.1-0.2 ~GeV^{2}$ corresponding usually to the
integrated total cross section, the diffraction slope is less than at 
$t=0$ by $\sim ~1 ~GeV^{-2}$ \cite{piN}. Therefore in all cases we report
%
%
\beq
B(\gamma^{*}\rightarrow V,x_{eff},Q^{2})|_{t=0} = 
B(\gamma^{*}\rightarrow V,x_{eff},Q^{2}) + 1 ~GeV^{-2}\,\, ,
\label{eq:19}
\eeq
%
%
where $B(x_{eff},Q^{2})$ 
is the diffraction slope determined  experimentally
from the $t$- integrated total cross section.
Eq. (\ref{eq:19}) gives an uncertainty $< 10\%$ in the value of $B$
and can be reduced if more accurate data will be appeared.

Combining the Eqs. (\ref{eq:14}), (\ref{eq:18}) and
(\ref{eq:19}) one can obtain the expression for
the dipole diffraction slope $B(x_{eff},r_{B})$:
%
%
\beq
B(x_{eff},r_{B}) = 
\biggl [1+B(\gamma^{*}\rightarrow V,x_{eff},Q^{2})
\biggr ]\,.\sqrt{\frac{g_{T}^{2}+
\epsilon\frac{Q^{2}}{m_{V}^{2}}g_{L}^{2}}
{h_{T}^{2}+
\epsilon\frac{Q^{2}}{m_{V}^{2}}h_{L}^{2}}}\,.
\frac{\sigma(x_{eff},r_{S})}{\sigma(x_{eff},r_{B})}\,\, ,
\label{eq:20}
\eeq
%
%
where $B(\gamma^{*}\rightarrow V,x_{eff},Q^{2})$ 
is the diffraction slope at
energy $W \sim \sqrt{\frac{Q^{2}+m_{V}^{2}}{x_{eff}}}$ and
at $Q^{2}$ taken from the data.
In (\ref{eq:20}) the values of the dipole cross section
$\sigma(x_{eff},r_{S})$ and  $\sigma(x_{eff},r_{B})$ are
also taken from the same data as the diffraction slope
following the procedure of extraction according to Eqs.
(\ref{eq:14}) and (\ref{eq:15}).
In (\ref{eq:20}) $\epsilon$ is the longitudinal polarization
of the photon with the values taken from the corresponding
experimental publications.
One can see from (\ref{eq:20}) that the dipole diffraction
slope is mainly determined by the slope parameter
$B(\gamma^{*}\rightarrow V,x_{eff},Q^{2})$
obtained from the data, because other two terms
in r.h.s. of (\ref{eq:20}),
\begin{displaymath}
\sqrt{\frac{g_{T}^{2}+
\epsilon\frac{Q^{2}}{m_{V}^{2}}g_{L}^{2}}
{h_{T}^{2}+
\epsilon\frac{Q^{2}}{m_{V}^{2}}h_{L}^{2}}}
~~~
and
~~~
\frac{\sigma(x_{eff},r_{S})}{\sigma(x_{eff},r_{B})}
\end{displaymath}
have the residual smooth $Q^{2}$ (dipole size) behaviour.
It is connected with the smooth $Q^{2}$ behaviour of the coefficient
functions $g_{T}$, $g_{L}$, $h_{T}$, $h_{L}$ and with the fact
that the ratio of two dipole cross sections at different
scanning radii $r_{S}$ and $r_{B}$ depends weakly on $r_{S}$. 
%
%
%
  \begin{figure}[tbh]
  \includegraphics{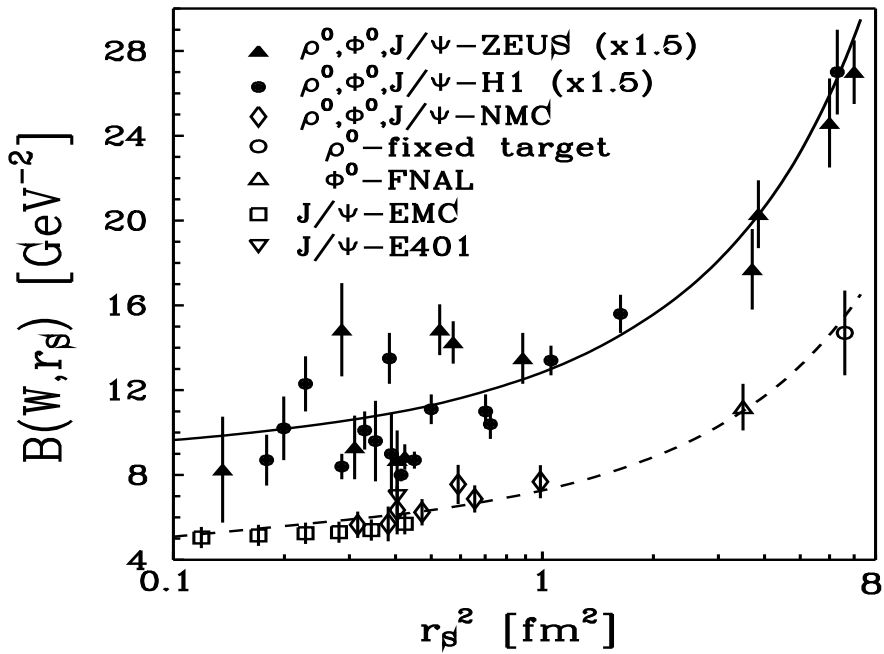}
  \begin{center}
  \vspace{8cm}
  \parbox{13cm}
  {\caption[Delta]
  {
   The color dipole size dependence
   of the dipole diffraction slope extracted from the 
   data on photoproduction and electroproduction of vector mesons:
   the low energy data on real photoproduction of $\rho^{0}$
   \cite{Rholownu},
   the low energy data on real photoproduction of $\Phi^{0}$
   \cite{Philownuold},
   the FNAL data on $\Phi^{0}$ production \cite{Philownu},
   the NMC data on $\Phi^{0}$ and $\rho^{0}$ production \cite{NMCfirho},
   the E401 data on $J/\Psi$ production \cite{E401Psi},
   the EMC data on $J/\Psi$ production \cite{EMCPsi,EMCPsiQ2},
   the NMC data on $J/\Psi$ production \cite{NMCPsi},
   the ZEUS data on $\rho^{0}$ production 
   \cite{ZEUSrho94,ZEUSrho95,ZEUSrho95Q2,ZEUSrho97,ZEUSrho98,ZEUSrho99Q2},
   the ZEUS data on $\Phi^{0}$ production \cite{ZEUSphi96},
   the ZEUS data on $J/\Psi$ production \cite{ZEUSPsi95,ZEUSPsi97},
   the H1 data on $\rho^{0}$ production
   \cite{H1rho96,H1rho96Q2,H1rho99Q2}
   the H1 data on $\Phi^{0}$ production
   \cite{H1phi97Q2}
   and H1 data on $J/\Psi$ production
   \cite{H1Psi94,H1Psi96Q2,H1Psi96,H1Psi99Q2,H1Psi00}.
   The dashed and solid curve show the dipole diffraction slope 
   of the model \cite{NZZslope,NNPZZ96bfkl} evaluated for the c.m.s.
   energy $W=15~GeV$ and $W=70~GeV$, respectively.
   The data points at HERA energies and the corresponding solid curve
   are multiplied by the factor 1.5
  }
  \label{event2}}
  \end{center}
  \end{figure}
%
%
%

The result of the above described analysis is depicted in
Fig.2, which shows the dipole size dependence of the dipole
diffraction slope extracted from the low energy and HERA data
of photoproduction and electroproduction of vector mesons.
The error bars shown here correspond to the error bars
in the measured production cross sections and diffraction
slopes as cited in the experimental papers.
Because the data on vector meson production fall into
two broad categories we present the procedure of extraction
for two energy ranges:
the center of mass energy $W\sim ~(10-15)~ GeV$ 
corresponding to fixed target data 
and $W\sim ~(70-150)~ GeV$ reflecting the HERA collider data. 
The color dipole cross section and the color dipole diffraction
slope is flavor blind, there is only kinematical dependence
on the vector meson through the definition of $x_{eff}$
(see Eq. (\ref{eq:8})).
However, because of the scanning phenomenon the comparison
of reactions with production of different vector mesons
at the same value of the scanning radius $r_{S}$
(at the same value of $Q^{2}+m_{V}^{2}$) leads approximately
to the same corresponding values of $x_{eff}$ at the fixed 
energy $\nu$.  
Thus, one expect that the extraction procedure (\ref{eq:20}),
(\ref{eq:14}) applied to the different vector mesons will
lead to the same value of $B(x_{eff},r_{S})$ at the same
value of the scanning radius.
This situation is similar to that for the color dipole cross
section. Again, the data show the decrease of the dipole
diffraction cone towards small dipole size $r_{S}$. 
It reflexes the contribution of the geometrical term $\propto r^2$
to the diffraction slope 
\cite{NZZslope,NNPZZ96bfkl}.
A comparison of the low energy fixed-target and HERA data
on real photoproduction and electroproduction of vector
mesons is not in contradiction with the conclusion about a substantial
shrinkage of the dipole diffraction cone coming from the gBFKL
phenomenology \cite{NZZslope,NNPZZ96bfkl}.
The corresponding effective shrinkage rate $\alpha'_{eff}$
for $\rho^{0}$ photoproduction is about $0.25 ~GeV^{-2}$
at the energy range of fixed target experiments and
slightly decreases to the value of about 0.2 GeV$^{-2}$
at HERA energy.
For electroproduction of charmonium and
electroproduction of $\rho^{0}$ when the 
scanning radius $\sim 0.2-0.3\,$ fm, the corresponding $\alpha'_{eff}\sim
0.15 ~GeV^{-2}$ at HERA energies 
in accordance with the gBFKL phenomenology
of a subasymptotic energy dependence of the diffraction slope
\cite{NZZslope,NNPZZ96bfkl}.
In another words, the shrinkage of the diffraction slope with
energy is weaker at smaller dipole sizes.
However, large error bars of the diffraction slope data
does not allow to see clearly the shrinkage of the slope
parameter with energy. Some evidence of that shrinkage is
seen only at large dipole sizes $r_{S}\gsim 1\,$ fm, where
the rise with energy of the slope is predicted to be more
substantial. 

The above determination of the color dipole diffraction slope
from the data
is rather crude for the following reasons
reported particularly in the paper \cite{NNPZdipole}:\\
{\bf i)}
The vector meson production data from the EMC collaboration
is known to have been plagued by a background from the
inelastic process $\gamma^{*}p\rightarrow VX$.
Especially at large $Q^{2}$ it could lead to enhancement
of the production cross section whereas the diffraction
slope could have been underestimated.
In the recent NMC data \cite{NMCfirho} a special care has been
taken to eliminate an inelastic background and 
the values of $B(x_{eff},r_{S})$ from the
NMC data are consistent within the experimental error bars. \\
{\bf ii)}
There are uncertainties connected with extrapolation
of the differential cross section down to $t=0$. Due to
the curvature of the diffraction cone the forward
production cross section can be underestimated. \\
{\bf iii)}
There is also conservative $\lsim 15\%$ theoretical
inaccuracy of the extraction procedure connected
with the variation of the coefficient functions 
$g_{T}$, $g_{L}$, $h_{T}$ and $h_{L}$ as a function of $Q^{2}$
from small ($W=15 ~GeV$) to large ($W=150 ~GeV$) energy
(see also paper \cite{NNPZdipole}).
\\
{\bf iiii)}
There is residual uncertainty connected with the wave function
of light vector mesons.

To summarize, within the above stated uncertainties of the
simple extraction procedure and the experimental errors
bars, there is a consistency between the dipole diffraction
slope determined from $\rho^{0}$, $\Phi^{0}$ and $J/\Psi$ 
production data. This is the first direct determination
of the dipole diffraction slope and the main conclusions
are not affected by the above cited uncertainties.

Fig.2 shows also the dipole diffraction slope from the
gBFKL analysis \cite{NZZslope,NNPZZ96bfkl}, which gives an unified
description of the data on vector meson diffractive production
and on the proton structure function.

{\bf Conclusions.} \\
We present the first determination of the dipole diffraction
slope from the data on diffractive photo- and electroproduction
of vector mesons. The dipole diffraction slope has been
evaluated at the dipole size down to $r_{S}\sim 0.35\,$ fm
and the decrease of $B(x_{eff},r_{S})$ towards small $r_{S}$
is in accordance with $r_{S}^{2}$ behaviour of the
diffraction slope coming from the gBFKL phenomenology.
Because of large error bars of the data,
we found only an evidence of a shrinkage 
of $B(x_{eff},r_{S})$
with energy from the fixed-target, $W\sim 10-15~GeV$
up to collider HERA energy range, $W\sim 70-150~GeV$.
This shrinkage is weaker for smaller dipole sizes.
The found pattern of dipole size and energy 
dependence of the dipole diffraction slope 
is consistent with the flavor independence
and with expectations from the gBFKL dynamics.

\end{document}